\documentclass[12pt]{article}
\usepackage{amssymb}
\usepackage{graphicx}

\ifx\pdfoutput\undefined
    \relax
\else
    \usepackage{epstopdf}
\fi
\makeatletter
\def\numberbysection{\@addtoreset{equation}{section}
 	\def\theequation{\thesection.\arabic{equation}}}
\makeatother

\numberbysection

\newcommand{\be}{\begin{eqnarray}}
\newcommand{\ee}{\end{eqnarray}}
\newcommand{\non}{\nonumber}

\newcommand{\sgn}{\mathop{\rm sgn}\nolimits}

\def\del{\bf\nabla}

\newcommand{\beq}{\begin{equation}}
\newcommand{\eeq}{\end{equation}}
\newcommand{\bea}{\begin{eqnarray*}}
\newcommand{\eea}{\end{eqnarray*}}
\newcommand{\beqa}{\begin{eqnarray}}
\newcommand{\eeqa}{\end{eqnarray}}

\begin{document}

\begin{titlepage}
\strut\hfill
\vspace{.5in}
\begin{center}

\LARGE Infinite degeneracy of Landau levels from the\\
\LARGE Euclidean symmetry with central extension revisited\\[1.0in]
\large Rajan Murgan\footnote{email:rmurgan@svsu.edu}\\[0.8in]
\large Department of Physics,\\ 
\large Saginaw Valley State University,\\ 
\large 7400 Bay Road University Center,  MI 48710 USA\\

\end{center}

\vspace{.5in}

\begin{abstract}

The planar Landau system which describes the quantum mechanical motion of a charged particle in a plane with a uniform magnetic field perpendicular to the plane, is explored within pedagogical settings aimed at the beginning graduate level. The system is known to possess the Euclidean symmetry in two dimensions with central extension $\bar{E}(2)$. In this paper, we revisit the well-known energy eigenvalues of the system, known as the Landau levels, by exploiting the related $\bar{e}(2)$ symmetry algebra. Specifically, we utilize the Casimir operator and the commutation relations of the generators of the $\bar{E}(2)$ group. More importantly, an algebraic formalism on this topic based on Schwinger's oscillator model of angular momentum is also presented. The dimensions of irreducible representations of the $\bar{E}(2)$ group and their implications on the degeneracy of Landau levels is discussed.

\end{abstract}
\end{titlepage}

\setcounter{footnote}{0}

\section{Introduction}\label{sec:intro}

Symmetries in quantum systems can provide great insights into their various properties. One such insight is on the nature of degeneracy of a system's energy levels. Examples of such quantum systems are the hydrogen atom \cite{Pauli, Fok} and higher dimensional quantum harmonic oscillators \cite{JH}. Another example is the planar Landau system, the subject of interest of this paper. This system refers to a quantum mechanical model of a charged particle moving in a plane with a perpendicular uniform magnetic field \cite{Landau1}. The Landau system has found profound applications in solid state physics and condensed matter physics. It plays important role as a fundamental model to understand many physical processes. Its energy eigenvalues, known in literature as Landau levels, have been a subject of intense studies in investigating electronic properties of materials such as the oscillations of electric resistivity and magnetic susceptibility as a function of the applied magnetic field. These are the Shubnikov-de Haas effect \cite{SchHaas1, SchHaas2} and de Haas-van Alphen effect \cite{deHaasAlphen} respectively. (Integer) quantum Hall effect is another indispensable example in this regard \cite{Klitzing}. All these critically reflect the importance of the Landau system in physics. The structure of Landau levels such as their degeneracy, is closely related to the system's symmetry. Thus, it is important that the subject of the system's symmetry group be introduced and adequately stressed especially within pedagogical realm.  

Planar Landau system is known to possess a dynamical symmetry represented by the Euclidean group in two dimensions with central extension, denoted hereafter by $\bar{E}(2)$. While this is a well-known fact in literature (see \cite{Neg, Jooh} and references therein), a coherent view on planar Landau system  and its symmetry is still lacking and often excluded from the discussion in standard graduate courses in quantum mechanics. This is a striking fact given the comparable applications and importance of this system with other crucial models in physics such as the quantum Kepler model and the quantum harmonic oscillator that receive much more exposure in pedagogical settings. Moreover, an approach predominantly based on group-theoretic method receives minimal coverage and often ignored in these courses despite its potential and importance as a tool to better understand and appreciate the rich mathematical structure associated with the dynamical symmetry of the Landau system. It is also worth pointing out that the $\bar{E}(2)$ symmetry, being an inherent symmetry of the system, is not apparent from the Hamiltonian itself and requires a little bit of discussion and algebra to reveal its presence. This may prove beneficial to students who already have some familiarity with the topic of angular momentum, symmetry operations and Lie groups such as the rotation group in three dimensions $SO(3)$.\footnote{$SO(3)$ is a Lie group of $3\times 3$ orthogonal matrices with determinant 1.}             

In this paper, we are motivated to address the above issues. Thus, our objective here is three-fold: Firstly, we intend to reintroduce the $\bar{E}(2)$ symmetry of the planar Landau system and revisit the associated centrally extended symmetry algebra (denoted hereafter by $\bar{e}(2)$). Secondly, we intend to use the $\bar{E}(2)$ group's Casimir operator to recover the well known Landau levels by performing the calculations separately in two sets of basis, which we shall show to be related to one another.\footnote{This is indeed given in \cite{Neg}, with the calculation carried out in one of the basis sets.} We will eventually use these basis sets to discuss the degeneracy of the Landau levels. By presenting the energy calculations in two basis sets, the relationship between the quantum numbers can be made more clear, which brings us to our foremost objective: The application of the techniques of Schwinger's oscillator model of angular momentum \cite{Schwinger, BV} within the framework of the Landau system. Schwinger's scheme features an interesting connection between the angular momentum algebra $so(3)$ and the algebra of two independent uncoupled oscillators \cite{SakuraiNep}. In this paper, we will exploit such a connection, but between the $\bar{e}(2)$ symmetry algebra (instead of $so(3)$\footnote{$so(3)$ refers to the algebra of the $SO(3)$ group, namely a set of commutation relations obeyed by the group's generators.}) and the oscillator algebra to proceed with the energy calculations. We find this approach to be very economical and yet to the best of our knowledge, we are unaware of such a presentation associated with the derivation of Landau levels in standard references. 

The paper is therefore structured as follows. In section 2, a brief review of the planar Landau system is given where the construction of its Hamiltonian is presented. In section 3, we review the essentials of the Euclidean group $E(2)$, and its central extension, the $\bar{E}(2)$ group. In section 4, the Hamiltonian of the planar Landau system is demonstrated to possess the symmetry $\bar{E}(2)$. The Casimir operator of this Lie group is written in terms of the Hamiltonian. This is then followed by the derivation of the well known Landau levels in two separate basis. While the calculation is based on Schwinger's oscillator model of angular momentum, it is developed here in terms of the $\bar{e}(2)$ algebra instead. All these are presented in section 5. Finally, we conclude this paper with a discussion in section 6.

\section{The planar Landau system}\label{sec:LS}

In this section, we review the Hamiltonian of the Landau system in $x$-$y$ plane. This refers to a quantum mechanical planar motion of a charge $q$ in a uniform magnetic field. The Hamiltonian $\mathcal{H}$ is given by\footnote{The classical mechanics version of the Hamiltonian is $\mathcal{H} = \frac{1}{2m}(\bold{P} - q\bold{A})^2$.}
\begin{equation}
\hat{\mathcal{H}} = \frac{1}{2m}(\hat{\bold{P}} - q\bold{A})^2\,, 
\label{Hamiltonian1}
\end{equation} 
where $m$ and $q$ refer to the mass and the charge of the particle respectively. The magnetic vector potential $\bold{A}$, is a function of the position operator $\hat{\bold{r}}$ which is defined as $\hat{\bold{r}} = \hat{x}\bold{n_{x}} + \hat{y}\bold{n_{y}} + \hat{z}\bold{n_{z}}$. $\{\bold{n_{x}}\,,\bold{n_{y}}\,,\bold{n_{z}}\}$ is the set of Cartesian unit vectors and $\{\hat{x}\,,\hat{y}\,,\hat{z}\}$ is the set of components of the three-dimensional position operator. It is important to note that the operator $\hat{\bold{P}}$ is the canonical momentum operator and not the kinetic momentum operator $m\hat{\bold{v}}$. They are related by
\begin{equation}
\hat{\bold{P}} = m\hat{\bold{v}} + q\bold{A}
\label{PpA}
\end{equation}  
Upon expanding the square in (\ref{Hamiltonian1}), one gets
\begin{equation}
\hat{\mathcal{H}} = \frac{1}{2m}(\hat{\bold{P}}^{2} - q(\bold{A}\cdot\hat{\bold{P}} + \hat{\bold{P}}\cdot\bold{A}) + q^{2}\bold{A}^{2})\,. 
\label{Hamiltonian2}
\end{equation} 
We recall here that $\bold{A}$ which is a function of position operator, does not in general commute with $\hat{\bold{P}}$. Indeed, we have
\begin{equation}
\bold{A}\cdot\hat{\bold{P}} - \hat{\bold{P}}\cdot\bold{A} = i\hbar \nabla\cdot\bold{A}\,,
\label{APcomm}
\end{equation}
where the following commutation rule for the (canonical) momentum operator with any function $g(\hat{\bold{r}})$, has been used (for example, see \cite{Landau2})\footnote{(\ref{Axcomm}) is a generalization of position-momentum commutation relation, $\hat{x}_{j}\hat{p}_{k} - \hat{p}_{k}\hat{x}_{j} = i\hbar\delta_{jk}\hat{\mathbb{I}}$}: 
\begin{equation}
g(\hat{\bold{r}})\hat{\bold{P}} - \hat{\bold{P}}g(\hat{\bold{r}}) = i\hbar \nabla g \,.
\label{Axcomm}
\end{equation}
In order to define $\bold{A}$ uniquely, we use the Coulomb/transverse gauge as the gauge-fixing condition and set $\del\cdot\bold{A} =$ 0. This choice also makes 
$\bold{A}$ and $\hat{\bold{P}}$ commute with each other. One possibility for $\bold{A}$ that satisfies the transverse gauge condition is the symmetric gauge, namely
\begin{equation}
\bold{A} = \frac{1}{2}\bold{B}\times \hat{\bold{r}} = -\frac{1}{2}\hat{y}B\bold{n_{x}} + \frac{1}{2}\hat{x}B\bold{n_{y}}\,.
\label{symmgauge}
\end{equation}
We have taken the magnetic field to be a constant and in the $z$-axis direction, $\bold{B} = B\bold{n_{z}}$. Next, we let $q = -|e|$ and bear in mind that $\hat{\bold{P}} = \hat{P_x}\bold{n_{x}} + \hat{P_y}\bold{n_{y}} + \hat{P_z}\bold{n_{z}}$. Using (\ref{APcomm}) and (\ref{symmgauge}) in (\ref{Hamiltonian2}), the Hamiltonian (in the symmetric gauge) assumes the following form,
\be
\hat{\mathcal{H}} &=& \frac{1}{2m}\hat{P_z}^2 + \frac{1}{2m}(\hat{P_x}^2 + \hat{P_y}^2) + \frac{m\omega^2}{2}(\hat{x}^2 + \hat{y}^2)
+\frac{|e|B}{2m}(\hat{x}\hat{P_y}-\hat{y}\hat{P_x})\non\\
&=& \hat{H}_{1} + \hat{H}_{2} + \hat{H}_{3}\,.
\label{Hamiltonianfinal}
\ee
$\omega = \frac{|eB|}{2m}$ is the Larmor frequency and 
\begin{equation}
\hat{H}_{1} = \frac{1}{2m}\hat{P_z}^2\,,\quad \hat{H}_{2} = \frac{1}{2m}(\hat{P_x}^2 + \hat{P_y}^2) + \frac{m\omega^2}{2}(\hat{x}^2 + \hat{y}^2)\,,\quad \hat{H}_{3} = \frac{|e|B}{2m}\hat{L_z}\,,  
\label{defs}
\end{equation} 
where $\hat{L_{z}} = \hat{x}\hat{P_y}-\hat{y}\hat{P_x}$. A few words on these terms are in order. $\hat{H}_{1}$ describes the charge's kinetic energy associated with its motion along the $z$-axis. From (\ref{symmgauge}), it is evident that the $z$-component of $\bold{A}$ is zero. Thus, it follows from (\ref{PpA}) that $\hat{P}_{z}$ is equal to the $z$-component of the kinetic momentum $m\hat{v}_{z}$. Since the eigenvalues of $\hat{P}_{z}$ can take all values, the velocity of the charge in the direction of $z$-axis can take any value as well. $\hat{H}_{2}$ resembles the Hamiltonian of the two-dimensional quantum harmonic oscillator and therefore possesses rotational symmetry about the $z$-axis. $\hat{H}_{3}$ is proportional to $\hat{L}_{z}$, a term that resembles the $z$-component of the angular momentum operator. However, as pointed out in \cite{Abers}, it is important to note that like $\hat{\bold{P}}$, $\hat{L_{z}}$ depends on the gauge and therefore is not an observable. This should be clear from (\ref{PpA}) that expresses $\hat{\bold{P}}$ in terms of $\bold{A}$, which is defined here in the transverse gauge. Since $\hat{L_{z}}$ is dependent on the components of $\hat{\bold{P}}$, it follows that $\hat{L_{z}}$ depends on the gauge as well and thus cannot represent the $z$-component of angular momentum operator which is indeed an observable. It is also worth noting that all three terms of the Hamiltonian (\ref{Hamiltonianfinal}) commute with each other. For a given eigenvalue of $\hat{H}_{1}$, the eigenvalues of the remaining terms known as the Landau levels are given by the following,
\begin{equation}
E_n = \hbar\omega(2n+1)\,,
\label{eigenv}
\end{equation}
where $n$ is a positive integer. The degeneracy of $E_n$ is not evident presently from (\ref{eigenv}). We shall address this crucial aspect of the planar Landau system in section 5 where we rederive (\ref{eigenv}) purely from symmetry considerations.  

\section{$E(2)$ and $\bar{E}(2)$ Lie Groups}\label{sec:E2}
In this section, we review important results on the Euclidean group in two-dimensions $E(2)$ and the Euclidean group in two-dimensions with central extension $\bar{E}(2)$.  
\subsection{$E(2)$ group}
$E(2)$ is a Lie group that is also known as the translation-rotation group in two dimensions. It defines 2 translation and 1 rotation symmetries.\footnote{Generally $E(n)$ defines $n$ translation symmetries and $\frac{n(n-1)}{2}$ rotation symmetries.} Mathematically, the group $E(2)$ is defined as a semidirect product of $T(2)$ and $O(2)$ Lie groups (written $T(2) \rtimes O(2)$), which are two-dimensional translation and orthogonal groups respectively (Refer to 
\cite{STberg} for a good description of this group.). The $E(2)$ group has a set of three Hermitian generators, $\{\hat{X}_{1}\,,\hat{X}_{2}\,,\hat{X}_{3}\}$. These generators satisfy a set of commutation relations known as the algebra of the group $e(2)$,
\be
\large[\hat{X}_{1},\hat{X}_{2}\large] = 0\,,\quad \large[\hat{X}_{3},\hat{X}_{1}\large] = i\hbar\hat{X}_{2}\,,\quad \large[\hat{X}_{3},\hat{X}_{2}\large] = -i\hbar\hat{X}_{1}\,.
\label{e2algebra}
\ee 
The Casimir operator (invariant operator) of the group, namely an operator that commutes with all the generators of the group is given by 
\be
\hat{C} = \hat{X}_{1}^2 + \hat{X}_{2}^2\,.
\label{CE2}
\ee
(\ref{e2algebra}) suggests that the set $\{\hat{X}_{1}\,,\hat{X}_{2}\,,\hat{X}_{3}\}$ represents the generators of translations in $x$-$y$ plane ($x$ and $y$ components of the (canonical) linear momentum operator) and rotations about the $z$-axis ($z$-component of the angular momentum operator) in that order (e.g., see \cite{Greiner}). The first commutator in (\ref{e2algebra}) reflects the usual commutative plane.  
\subsection{$\bar{E}(2)$ group}
$\bar{E}(2)$ is essentially $E(2)$ with central extension. Unlike the $E(2)$ group, the $\bar{E}(2)$ does not receive as much exposure in standard group theory references as the $E(2)$. It has a set of four Hermitian generators, namely $\{\hat{Y}_{1}\,,\hat{Y}_{2}\,,\hat{Y}_{3}\,,\hat{D}\}$ that includes a new central generator $\hat{D}$. This set is related to that of $\{\hat{X}_{1}\,,\hat{X}_{2}\,,\hat{X}_{3}\}$ by,
\be
\hat{Y}_i = \hat{X}_i + \hat{W}_i\,,\quad i=1\,,2\,,3\,,
\label{corresp}
\ee
As will be clear in the next section, $\hat{W}_{i}$ is an operator that is a function of the eigenvalue of $\hat{D}$ \cite{Neg}. The related algebra $\bar{e}(2)$ obeyed by the four generators are given by the following commutation relations,
\be
\large[\hat{Y}_{1},\hat{Y}_{2}\large] = i\hbar\hat{D}\,,\quad \large[\hat{Y}_{3},\hat{Y}_{1}\large] = i\hbar\hat{Y}_{2}\,,\quad \large[\hat{Y}_{3},\hat{Y}_{2}\large] = -i\hbar\hat{Y}_{1}\,.
\label{e2baralgebra}
\ee 
The generator $\hat{D}$ commutes with all the other generators of the $\bar{E}(2)$ group.\footnote{Despite this property of $\hat{D}$ and the definition of the Casimir operator, we stress here that $\hat{D}$ is not the Casimir operator of the group but simply another generator of the group which together with other generators of the group, form the Lie algebra $\bar{e}(2)$.} It will be clear shortly in the next section that the first commutator in (\ref{e2baralgebra}) indicates the presence of a non-commutative plane within the context of planar Landau system. The Casimir operator of the $\bar{E}(2)$ group is given by
\be
\hat{\bar{C}} = \hat{Y}_{1}^2 + \hat{Y}_{2}^2 + 2\hat{Y}_{3}\hat{D}\,.
\label{CE2bar}
\ee
One notes an obvious difference between (\ref{CE2}) and (\ref{CE2bar}) where in latter, the operator $\hat{D}$ that plays the role of the non-commutativity factor in (\ref{e2baralgebra}) is present. It will also be evident that within the context of Landau system, this operator will define the dynamic feature of the symmetry of the system, an essential factor in defining the degeneracy of Landau levels.

\section{$\bar{E}(2)$ group and the planar Landau system}
Symmetry and degeneracy in a system are related to each other. Symmetries are usually attributed to the presence of conserved observables. An observable of a system $\hat{G}$ is conserved if it commutes with the Hamiltonian of the system $\hat{\cal{H}}$, since $\frac{d\hat{G}}{dt} = \large[\hat{G},\hat{\cal{H}}\large]$, provided $\hat{G}$ does not explicitly depend on time, $t$. It is therefore instructive to find such conserved observables for the planar Landau system that would lead us to the symmetry of the system and therefore to the subject of degeneracy of Landau levels. Prior to that, in order to make the paper self-contained, it is essential to recall and review an important result regarding conserved observables, symmetry and degeneracy along the line given in \cite{Landau2}: The existence of a further conserved observable, which cannot be measured simultaneously with the others, leads to an additional degeneracy of the energy levels. A good discussion on this can also be found in \cite{Greiner}. Suppose that $\hat{K}$ and $\hat{G}$ are two conserved observables (they each commute with the Hamiltonian, $\hat{\cal{H}}$) that do not commute with each other,
\be
\Large[\hat{K},\hat{\cal{H}}\Large] &=& 0 \non\\
\Large[\hat{G},\hat{\cal{H}}\large] &=& 0 \non\\
\Large[\hat{K},\hat{G}\Large]&\neq0 & \,.
\label{kgH}
\ee
We denote $\{\vert\phi_{j}\rangle\}$ as the complete common set of eigenkets of $\hat{K}$ and $\hat{\cal{H}}$ (that is to say $\{\vert\phi_{j}\rangle\}$ represents the complete set of state vectors that simultaneously diagonalizes both $\hat{\cal{H}}$ and ${\hat{K}}$), namely
\be
\hat{\cal{H}}\vert\phi_{j}\rangle &=& E_{j}\vert\phi_{j}\rangle\non\\
\hat{K}\vert\phi_{j}\rangle &=& K_{j}\vert\phi_{j}\rangle\,,\quad j = 1\,,2\,,\ldots\/,.
\label{kheigenv}
\ee
$K_{j}$ and $E_{j}$ above represent the corresponding eigenvalues of $\hat{K}$ and $\hat{\cal{H}}$ respectively. Now, let $\vert\phi\rangle$ be any one of the eigenkets from the set $\{\vert\phi_{j}\rangle\}$ and $E$ being the corresponding energy eigenvalue. Since $\hat{K}$ and $\hat{G}$ do not commute, $\hat{G}\vert\phi\rangle$ does not coincide with $\vert\phi\rangle$. If they coincide, this would imply that $\hat{G}$ also has a definite value in the state $\vert\phi\rangle$ which would then imply simultaneous measurement of $K$ and $G$. On the other hand, since $\large[\hat{G},\hat{\cal{H}}\large] = 0$,
\begin{equation}
\hat{\cal{H}}(\hat{G}\vert\phi\rangle)=\hat{G}\hat{\cal{H}}\vert\phi\rangle = \hat{G}E\vert\phi\rangle=E(\hat{G}\vert\phi\rangle)
\label{kgH2}
\end{equation}
We then see that both the states $\vert\phi\rangle$ and $\hat{G}\vert\phi\rangle$ correspond to the same energy $E$, thus demonstrating degeneracy of the energy level.

\subsection{Conserved observables for Landau system}
Observables that are conserved in the planar Landau system are well-known. A clear account on these observables is given in \cite{Landau2, Abers}. While the presence of these observables are not apparent from the system's Hamiltonian (\ref{Hamiltonianfinal}), a good intuitive understanding of these observables can be obtained by revisiting the system's classical counterpart: In classical mechanics, the motion of the charge subjected to a constant magnetic field along the $z$-axis ($B\bold{n_z}$) is along a helix, with the Hamiltonian of the system given in footnote 4. The axis of the helix is along the $z$-axis and is fixed on coordinates $x = x_{o}\,,y = y_{o}$. Another way to think about this within the classical realm is by ignoring the drift due to the $z$-component of the constant velocity of the charge. The motion of the charge then is essentially in the $x$-$y$ plane, in a circle that is perpendicular to the magnetic field. Thus, $x_{o}$ and $y_{o}$ correspond to the $x$ and $y$ coordinates of the centre of the circle. The general solution of the classical equations of motion for this system is given by,
\be
x_o &=& x -\sgn(B) \frac{mv_{y}}{2m\omega} = \frac{1}{2}(x -\sgn(B)\frac{P_y}{m\omega})\,,\non\\
y_o &=& y +\sgn(B) \frac{mv_{x}}{2m\omega} = \frac{1}{2}(y +\sgn(B)\frac{P_x}{m\omega})\,,  
\label{xoyoclassical}
\ee
where $\sgn(B)$ is defined as 
\be
\sgn(B) = \left\{ \begin{array}{c@{\quad : \quad} l}
+ & B > 0 \\
- & B < 0\\
\end{array} \right. \,,
\label{sgnB}
\ee 
and the relation between the kinetic momentum $m\vec{v}$ and canonical momentum $\vec{P}$, namely $mv_i=P_i+|e|A_i$ is used. $A_{i}$ represents a component of the magnetic vector potential in symmetric gauge, $\vec{A} = -\frac{1}{2}yB\bold{n_{x}} + \frac{1}{2}xB\bold{n_{y}}$.

In the quantum description of the problem, the operators $\hat{x}_{o}$ and $\hat{y}_{o}$ are the conserved observables, thus commuting with the Hamiltonian of the system. For now, we simply state without giving the proof that $\hat{x}_{o}$ and $\hat{y}_{o}$ do not comute with each other. The proof for this will follow after we define some crucial operators. 

$\hat{H}_{2}$ in (\ref{Hamiltonianfinal}) that is essentially the Hamiltonian of the two-dimensional quantum harmonic oscillator suggests we start by defining a familiar set of annihilation operators $\{\hat{a}_{x}\,,\hat{a}_{y}\}$ and creation operators $\{\hat{a}^{\dagger}_{x}\,,\hat{a}^{\dagger}_{y}\}$ used in the quantum harmonic oscillator formalism:
\be
\hat{a}_{x}&=&\sqrt{\frac{m\omega}{2\hbar}}(\hat{x} + \frac{i}{m\omega}\hat{P}_{x})\,,\quad \hat{a}^{\dagger}_{x} = \sqrt{\frac{m\omega}{2\hbar}}(\hat{x} - \frac{i}{m\omega}\hat{P}_{x})\,,\non\\
\hat{a}_{y}&=&\sqrt{\frac{m\omega}{2\hbar}}(\hat{y} + \frac{i}{m\omega}\hat{P}_{y})\,,\quad \hat{a}^{\dagger}_{y} = \sqrt{\frac{m\omega}{2\hbar}}(\hat{y} - \frac{i}{m\omega}\hat{P}_{y})\,,
\label{axyhats}
\ee  
with the following non-vanishing commutators,
\be
\Large[\hat{a}_{x},\hat{a}^{\dagger}_{x}\Large] = \hat{\mathbb{I}}\,,\quad \Large[\hat{a}_{y},\hat{a}^{\dagger}_{y}\Large] = \hat{\mathbb{I}}\,.
\label{axycomms}
\ee
In terms of the above annihilation and creation operators, $\hat{H}_2$ and $\hat{H}_3$ in (\ref{Hamiltonianfinal}) become
\be
\hat{H}_{2} = \hbar\omega(\hat{a}^{\dagger}_{x}\hat{a}_{x} + \hat{a}^{\dagger}_{y}\hat{a}_{y} + \hat{\mathbb{I}})\,,\quad \hat{H}_{3} = i\hbar\frac{|e|B}{2m}(\hat{a}_{x}\hat{a}^{\dagger}_{y} - \hat{a}_{y}\hat{a}^{\dagger}_{x})\,.
\label{h1h2axay}
\ee
Terms in $\hat{H}_{3}$ can be neatly decoupled into mutually commuting terms by rewriting it in terms of a new set of annihilation $\{\hat{a}_{-}\,,\hat{a}_{+}\}$ and creation 
$\{\hat{a}^{\dagger}_{-}\,,\hat{a}^{\dagger}_{+}\}$ operators \cite{Abers},
\be
\hat{a}_{+} &=& \frac{1}{\sqrt{2}}(\hat{a}_{x} - i\hat{a}_{y})\,,\quad \hat{a}^{\dagger}_{+} = \frac{1}{\sqrt{2}}(\hat{a}^{\dagger}_{x} + i\hat{a}^{\dagger}_{y})\non\\
\hat{a}_{-} &=& \frac{1}{\sqrt{2}}(\hat{a}_{x} + i\hat{a}_{y})\,,\quad \hat{a}^{\dagger}_{-} = \frac{1}{\sqrt{2}}(\hat{a}^{\dagger}_{x} - i\hat{a}^{\dagger}_{y})\,,
\label{apam}
\ee
which then makes (\ref{Hamiltonianfinal}) to assume the following form,
\be
\hat{\mathcal{H}} = \frac{\hat{p}^{2}_{z}}{2m} + \hbar\omega(\hat{a}^{\dagger}_{+}\hat{a}_{+} + \hat{a}^{\dagger}_{-}\hat{a}_{-} + \hat{\mathbb{I}}) + \frac{|e|B}{2m}\hbar(\hat{a}^{\dagger}_{+}\hat{a}_{+} - \hat{a}^{\dagger}_{-}\hat{a}_{-})\,.
\label{hapam}
\ee
$\hat{H}_3$ in (\ref{defs}) then implies,
\be
\hat{L}_z = \hbar(\hat{a}^{\dagger}_{+}\hat{a}_{+} - \hat{a}^{\dagger}_{-}\hat{a}_{-})\,,
\label{lzapm}
\ee 
an expression that is indispensable for the calculation of Landau levels in section 5. Following \cite{SakuraiNep}, operators $\{\hat{a}_{+}\,,\hat{a}^{\dagger}_{+}\}$ are defined as the annihilation and creation operators for the plus-type oscillators and operators $\{\hat{a}_{-}\,,\hat{a}^{\dagger}_{-}\}$ are defined as the annihilation and creation operators for the minus-type oscillators. Commutation relations among $\{\hat{a}_{\pm}\,,\hat{a}^{\dagger}_{\pm}\}$ can be obtained from (\ref{axycomms}) and (\ref{apam}). They are given below,
\be
\Large[\hat{a}_{+},\hat{a}^{\dagger}_{+}\Large] = \hat{\mathbb{I}}\,,\quad \Large[\hat{a}_{-},\hat{a}^{\dagger}_{-}\Large] = \hat{\mathbb{I}}\,,\quad 
\Large[\hat{a}_{+},\hat{a}^{\dagger}_{-}\Large] = \Large[\hat{a}_{-},\hat{a}^{\dagger}_{+}\Large] = \Large[\hat{a}_{-},\hat{a}_{+}\Large] = 0\,.
\label{apmcomms}
\ee 
Clearly, commutators of operators of different oscillators commute (the plus and minus oscillators are uncoupled). 
Our motivation to introduce (\ref{apam}) here is as follows. We would like to exploit the relationship between the $\bar{e}(2)$ algebra and the algebra of two independent oscillators to study the planar Landau system and the Landau levels where the set (\ref{apam}) is proven to be immensely useful. This is based on  Schwinger's oscillator model of angular momentum \cite{Schwinger, BV}. Schwinger's formalism relates in a very interesting way the $so(3)$ algebra of angular momentum and the algebra of two independent (uncoupled) oscillators utilizing the set (\ref{apam}). 

Now, we are ready to define the two non-commuting conserved observables mentioned earlier. They can be conveniently expressed in terms of the annihilation-creation operators defined in (\ref{apam})\cite{Abers}:
\be
\hat{x}_{o} &=& \frac{1}{2}\sqrt{\frac{\hbar}{m\omega}}(\hat{a}^{\dagger}_{-} + \hat{a}_{-})\,,\quad \hat{y}_{o} = \frac{i}{2}\sqrt{\frac{\hbar}{m\omega}}(\hat{a}^{\dagger}_{-} - \hat{a}_{-})\,,\quad B > 0\,,\non\\
\hat{x}_{o} &=& \frac{1}{2}\sqrt{\frac{\hbar}{m\omega}}(\hat{a}^{\dagger}_{+} + \hat{a}_{+})\,,\quad \hat{y}_{o} = -\frac{i}{2}\sqrt{\frac{\hbar}{m\omega}}(\hat{a}^{\dagger}_{+} - \hat{a}_{+})\,,\quad B < 0\,,
\label{xoyo}
\ee
The sign of $B$ dictates the set of operators to use. Using (\ref{axyhats}) and (\ref{apam}), $\hat{x}_{o}$ and $\hat{y}_{o}$ can be expressed in terms of $\hat{x}\,,\hat{y}\,,\hat{P}_{x}$ and $\hat{P}_{y}$,\be
\hat{x}_{o} = \frac{1}{2}(\hat{x} -\sgn(B) \frac{\hat{P}_{y}}{m\omega})\,,\quad \hat{y}_{o} = \frac{1}{2}(\hat{y} +\sgn(B) \frac{\hat{P}_{x}}{m\omega})\,.
\label{xoyoxypxpy}
\ee
(\ref{xoyoxypxpy}) is the quantum version of (\ref{xoyoclassical}). (\ref{Hamiltonianfinal}) and (\ref{xoyoxypxpy}) in conjunction with the usual position-momentum commutation relation yield the following results for the planar Landau system which are crucial in reflecting degeneracy of the Landau levels,
\be
\Large[\hat{x}_{o},\hat{\mathcal{H}}\Large] = 0\,,\quad \Large[\hat{y}_{o},\hat{\mathcal{H}}\Large] = 0\,,\quad \Large[\hat{x}_{o},\hat{y}_{o}\Large] = \sgn(B)\frac{i\hbar}{|eB|}\hat{\mathbb{I}}\,.
\label{degencomm}
\ee 
The definition $\omega = \frac{|eB|}{2m}$ has been used in evaluating the commutator of $\hat{x}_{o}$ and $\hat{y}_{o}$. It is clear from the first two commutators in (\ref{degencomm}) that $\hat{x}_o$ and $\hat{y}_o$ are conserved. Two important points now follow: Firstly, (\ref{degencomm}) satisfies the conditions in (\ref{kgH}). Thus, the argument that follows (\ref{kgH}) can be applied to (\ref{degencomm}), which essentially confirms degeneracy in Landau levels. Next, the last commutator in (\ref{degencomm}) reflects the physical realization of non-commutative plane in planar Landau system. This will be elaborated further below in terms of the generators of the $\bar{E}(2)$ group.

\subsection{Generators of the $\bar{E}(2)$ group}
Here we shall present the four generators that obey (\ref{e2baralgebra}), two of them in terms of the conserved observables of the planar Landau system, $\hat{x}_{o}$ and $\hat{y}_{o}$. They commute with the system's Hamiltonian (\ref{Hamiltonianfinal}), which reflects the symmetry of the system. These generators are as follows,
\be
\hat{J}_{1} = -\sqrt{|e|}B\hat{y}_o\,,\quad \hat{J}_{2} = \sqrt{|e|}B\hat{x}_o\,,\quad \hat{J}_{3} = \hat{M}_z = \hbar(\hat{a}^{\dagger}_{+}\hat{a}_{+}-\hat{a}^{\dagger}_{-}\hat{a}_{-})\,,\quad \hat{B} = B\hat{\mathbb{I}}\,,
\label{e2barxoyogen}
\ee    
that obey the following commutation rules:
\be
\large[\hat{J}_{1},\hat{J}_{2}\large] = i\hbar B\hat{\mathbb{I}}\,,\quad \large[\hat{J}_{3},\hat{J}_{1}\large] = i\hbar\hat{J}_{2}\,,\quad \large[\hat{J}_{3},\hat{J}_{2}\large] = -i\hbar\hat{J}_{1}\,.
\label{algeb}
\ee
We point out that while the third observable $\hat{M}_{z}$ looks identical to the $\hat{L}_{z}$ operator given in (\ref{lzapm}), as stressed earlier, $\hat{L}_{z}$ is a term that appears in (\ref{Hamiltonianfinal}) due to our choice of the gauge-fixing condition. $\hat{M}_{z}$ is an observable inherent to the system and indeed is the $z$-component of the angular momentum. The fourth generator $\hat{B}$ is simply the constant magnetic field of the system. We immediately note that (\ref{algeb}) coincides with the $\bar{e}(2)$ algebra given by (\ref{e2baralgebra}). Moreover, using (\ref{xoyoxypxpy}), (\ref{e2barxoyogen}) assumes the structure (\ref{corresp}),
\be
\hat{J}_{1} &=& -\frac{1}{\sqrt{|e|}}\hat{P}_{x} + \hat{W}_{1}\,,\quad \hat{W}_{1} = -\frac{1}{2}\sqrt{|e|}B\hat{y}\non\\
\hat{J}_{2} &=& -\frac{1}{\sqrt{|e|}}\hat{P}_{y} + \hat{W}_{2}\,,\quad \hat{W}_{2} = \frac{1}{2}\sqrt{|e|}B\hat{x}\non\\
\hat{J}_{3} &=& \hat{M}_{z}\,,\quad \hat{W}_{3} = 0\,.
\label{yxpw}
\ee 
The following correspondence can then be made:
\be
\{\hat{Y}_{1}\,,\hat{Y}_{2}\,,\hat{Y}_{3}\,,\hat{D}\}\to\{\hat{J}_{1}\,,\hat{J}_{2}\,,\hat{J}_{3}\,,B\hat{\mathbb{I}}\}\,,\quad \{\hat{X}_{1}\,,\hat{X}_{2}\,,\hat{X}_{3}\}\to\{-\frac{1}{\sqrt{|e|}}\hat{P}_{x}\,,-\frac{1}{\sqrt{|e|}}\hat{P}_{y}\,,\hat{M}_{z}\}\,,
\label{settoset}
\ee
We note that even with the inclusion of the scalar term $-\frac{1}{\sqrt{|e|}}$ in $\hat{P}_x$ and $\hat{P}_y$, the commutation relations among $\{-\frac{1}{\sqrt{|e|}}\hat{P}_{x}\,,-\frac{1}{\sqrt{|e|}}\hat{P}_{y}\,,\hat{M}_{z}\}$ are still given by (\ref{e2algebra}). As stated in section 3.2, (\ref{yxpw}) indicates that the $\hat{W}_{1}\,,\hat{W}_{2}$ are indeed functions of eigenvalue of $\hat{D}$, which (\ref{settoset}) simply shows to be $B$. As discussed in section 3.2, the commutator of $\hat{J}_1$ and $\hat{J}_2$ reflects the physical realization of the non-commutative plane in the planar Landau system. The magnetic field $B$, plays the role of the non-commutativity factor. Additionally, due to the important role that $B$, a dynamical factor of the system, plays in defining the symmetry of the planar Landau system, this symmetry can be termed as the dynamical symmetry of the system.\footnote{See\cite{Greiner} for a good discussion on dynamical symmetries in quantum systems.}Further, using (\ref{hapam}), (\ref{degencomm}) and (\ref{e2barxoyogen}) one deduces the following,
\be
\large[\hat{J}_{k},\hat{\mathcal{H}}\large] &=& 0\,,\quad k = 1\,,2\,,3\,.\non\\
\large[\hat{B},\hat{\mathcal{H}}\large] &=& 0\,.
\label{Hsymm}
\ee
One can then conclude that the Hamiltonian (\ref{Hamiltonianfinal}) possesses the symmetry associated with $\bar{e}(2)$ algebra (\ref{algeb}), namely the $\bar{E}(2)$ symmetry. 

\section{Landau levels and $\bar{E}(2)$ Casimir operator}\label{sec:Lls}
In this section, we revisit the role of the Casimir operator of the $\bar{E}(2)$ Lie group in the derivation of the Landau levels. Prior to that, we shall recast (\ref{CE2bar}) in terms of the Hamiltonian $\hat{H}_{2} + \hat{H}_{3}$.\footnote{In literature such as \cite{Neg,Jooh}, the Casimir operator is taken to be the Hamiltonian. We have decided to maintain the difference between them despite the fact that one is indeed simply proportional to the other.} More importantly, the main result of this paper will be presented: Utilization of the relationship between the $\bar{e}(2)$ algebra and the algebra of two independent oscillators in the derivation of the Landau levels, a method based on Schwinger's oscillator model of angular momentum.

Using the first correspondence in (\ref{settoset}), we first rewrite the Casimir operator of the $\bar{E}(2)$ group given by (\ref{CE2bar}), 
\be
\hat{\bar{C}} &=& \hat{J}^{2}_{1} + \hat{J}^{2}_{2} + 2\hat{J}_{3}\hat{B}\,,\non\\
              &=& |e|B^{2}\hat{y_o}^{2} + |e|B^{2}\hat{x_o}^{2} + 2\hat{M}_{z}B\,,
\label{ce2barnew}
\ee
in terms of the annihilation and creation operators $\{\hat{a}_{\pm}\,,\hat{a}^{\dagger}_{\pm}\}$ introduced in (\ref{apam}). This, upon using (\ref{xoyo}) and (\ref{e2barxoyogen}) becomes,
\be
\hat{\bar{C}} &=& \hbar|B|(2\hat{a}^{\dagger}_{+}\hat{a}_{+} + \hat{\mathbb{I}})\,,\quad B > 0\non\\
\hat{\bar{C}} &=& \hbar|B|(2\hat{a}^{\dagger}_{-}\hat{a}_{-} + \hat{\mathbb{I}})\,,\quad B < 0\,.
\label{CsgnB}
\ee       
which can be expressed in terms of the sum of the last two terms in (\ref{hapam}), namely $\hat{H}_{23} = \hat{H}_{2}+\hat{H}_{3}$ as
\be
\hat{\bar{C}} = \frac{2m}{|e|}\hat{H}_{23}\,.
\label{H23}
\ee

\subsection{Relation between the $\bar{e}(2)$ algebra and the oscillator algebra}
In order to demonstrate the connection between the two algebras, we utilize a set of raising and lowering operators $\{\hat{J}_{+}\,,\hat{J}_{-}\}$ constructed from two of the $\bar{E}(2)$ generators,
\be
\hat{J}_{\pm} = \frac{1}{\sqrt{2}}(\hat{J}_{1} \pm i\hat{J}_{2})\,.
\label{jpm}
\ee 
$\hat{J}_{\pm}$, together with the remaining generators of the group, namely the set $\{\hat{J}_{+}\,,\hat{J}_{-}\,,\hat{J}_{3}\,,\hat{B}\}$ proves to be very useful in the calculation of Landau levels. In terms of the raising and lowering operators (\ref{jpm}), the $\bar{e}(2)$ algebra (\ref{algeb}) can be rewritten as,
\be
\large[\hat{J}_{+},\hat{J}_{-}\large] = \hbar B\hat{\mathbb{I}}\,,\quad \large[\hat{J}_{3},\hat{J}_{+}\large] = \hbar\hat{J}_{+}\,,\quad \large[\hat{J}_{3},\hat{J}_{-}\large] = -\hbar\hat{J}_{-}\,.
\label{e2bcartanalgeb}
\ee
The algebra of two independent simple harmonic oscillators can be written down using the plus-type and minus-type annihilation and creation operators defined in (\ref{apam}). Part of the commutation relations are already given in (\ref{apmcomms}). Two operators that play instrumental role in oscillator algebra are the number operators:
\be
\hat{N}_{\pm} = \hat{a}^{\dagger}_{\pm}\hat{a}_{\pm}\,,
\label{numberpm}
\ee    
which can be readily seen to commute with each other. Using (\ref{apmcomms}), they can be shown to form the following commutation rules with the plus and minus-type creation and annihilation operators,
\be
\large[\hat{N}_{+},\hat{a}_{+}\large] = -\hat{a}_{+}\,,\quad\large[\hat{N}_{-},\hat{a}_{-}\large] = -\hat{a}_{-}\,,\quad\large[\hat{N}_{+},\hat{a}^{\dagger}_{+}\large] = \hat{a}^{\dagger}_{+}\,,\quad\large[\hat{N}_{-},\hat{a}^{\dagger}_{-}\large] = \hat{a}^{\dagger}_{-}\,.
\label{morecom}
\ee
Furthermore, one can use (\ref{xoyo}) and (\ref{e2barxoyogen}) to rewrite the $\hat{J}_{\pm}$ operators in terms of $\{\hat{a}_{\pm}\,,\hat{a}^{\dagger}_{\pm}\}$ operators that form the oscillator algebra. They assume the following form,
\be
\hat{J}_{+} &=& i\sqrt{\hbar B}\hat{a}_{-}\,,\quad \hat{J}_{-} = -i\sqrt{\hbar B}\hat{a}^{\dagger}_{-}\,,\quad B > 0\non\\
\hat{J}_{+} &=& -i\sqrt{\hbar |B|}\hat{a}^{\dagger}_{+}\,,\quad \hat{J}_{-} = i\sqrt{\hbar |B|}\hat{a}_{+}\,,\quad B < 0\,.
\label{oscie2}
\ee 
Together with the operator $\hat{J}_{3} = \hat{M}_{z} = \hbar(\hat{a}^{\dagger}_{+}\hat{a}_{+} - \hat{a}^{\dagger}_{-}\hat{a}_{-})$, one can show that these three operators which are constructed from the oscillator operators indeed satisfy the $\bar{e}(2)$ commutation relations given in (\ref{e2bcartanalgeb}). This demonstrates the connection between the oscillator algebra and $\bar{e}(2)$.\footnote{It is important to note here that in Schwinger's oscillator model of angular momentum, the raising and lowering operators are defined as $\hat{J}_{\pm} = \hbar\hat{a}^{\dagger}_{\pm}\hat{a}_{\mp}$. Together with $\hat{J}_{z} = \frac{\hbar}{2}(\hat{a}^{\dagger}_{+}\hat{a}_{+} - \hat{a}^{\dagger}_{-}\hat{a}_{-})$, these can be shown to satisfy the well-known angular-momentum commutation relations that form the $so(3)$ algebra\cite{SakuraiNep}.}

\subsection{Landau levels in $|l\,,m\,,B\rangle$ basis}
We shall first demonstrate Landau levels calculation as presented in \cite{Neg}. The basis used is formed by the set of state vectors $\{|l\,,m\,,B\rangle\}$ that serves as a common set of eigenkets for $\hat{\bar{C}}\,,\hat{M}_{z}$ and $\hat{B}$:
\be
\hat{\bar{C}}|l\,,m\,,B\rangle = C_{l}|l\,,m\,,B\rangle\,,\quad\hat{M}_{z}|l\,,m\,,B\rangle = m\hbar|l\,,m\,,B\rangle\,,\quad\hat{B}|l\,,m\,,B\rangle = B|l\,,m\,,B\rangle\,,
\label{lmb}
\ee
where $l$ is the quantum number associated with the eigenvalues of Casimir operator $\hat{\bar{C}}$ (and therefore also of the Hamiltonian $\hat{H}_{23}$),
$m$ is the quantum number associated with the eigenvalues of angular momentum $\hat{M}_{z}$ and $B$ is the magnetic field. Prior to proceeding with the calculation, an important point follows from (\ref{e2bcartanalgeb}): $\hat{J}_{-}$ lowers the eigenvalue of $z$-component of angular momentum by $\hbar$ and $\hat{J}_{+}$ raises the eigenvalue by $\hbar$. This implies the following:
\be
\hat{J}_{-}|l\,,m\,,B\rangle = \alpha|l\,,m-1\,,B\rangle\,,\quad \hat{J}_{+}|l\,,m\,,B\rangle = \beta|l\,,m+1\,,B\rangle\,,
\label{raiselowerkets}
\ee
where $\alpha$ and $\beta$ are dependent on $m$ and $l$. We shall determine these for $B<0$ and $B>0$ cases shortly. If $-l$ is taken to be the lowest $m$ value (for a given $l$), it follows that $|l,-l,B\rangle$ is the lowest state and therefore one concludes
\be
\hat{J}_{-}|l\,,-l\,,B\rangle = 0\,.
\label{jmin}
\ee  
Likewise, if $l$ is taken to be the highest $m$ value (for a given $l$), it follows that $|l,l,B\rangle$ is the highest state and thus 
\be
\hat{J}_{+}|l\,,l\,,B\rangle = 0\,.
\label{jmax}
\ee  
It is now imperative to rewrite the Casimir operator $\hat{\bar{C}}$ in terms of the raising and lowering operators. From (\ref{algeb}), (\ref{ce2barnew}) and (\ref{jpm}) one can recast the Casimir operator as follows,
\be
\hat{\bar{C}} = 2\hat{J}_{+}\hat{J}_{-} + 2\hat{M}_{z}B - B\hbar\hat{\mathbb{I}}\,,
\label{Cjm}
\ee
and 
\be
\hat{\bar{C}} = 2\hat{J}_{-}\hat{J}_{+} + 2\hat{M}_{z}B + B\hbar\hat{\mathbb{I}}\,.
\label{Cjp}
\ee
We first perform the calculation using the lowest state: Acting $\hat{\bar{C}}$ on the lowest state $|l\,,-l\,,B\rangle$ as follows,
\be
\hat{\bar{C}}|l\,,-l\,,B\rangle = 2\hat{J}_{+}\hat{J}_{-}|l\,,-l\,,B\rangle + 2\hat{M}_{z}B|l\,,-l\,,B\rangle - B\hbar\hat{\mathbb{I}}|l\,,-l\,,B\rangle\,,
\label{Conlow}
\ee 
and exploiting (\ref{jmin}), the following is obtained for the eigenvalues of the Casimir operator,
\be
C_l = -B\hbar(2l + 1)\,,\quad B < 0\,.
\label{Clow}
\ee
We equate this to eigenvalues of the Casimir operator obtained using the common set of eigenkets of the Hamiltonian $\hat{H}_{23}$ and $\hat{\bar{C}}$ (as suggested by(\ref{H23})), 
\be
\hat{\bar{C}}|E\rangle = \frac{2m}{|e|}\hat{H}_{23}|E\rangle = \frac{2m}{|e|}E_l|E\rangle\,,
\label{CH23}
\ee
where $E_{l}$ represents the eigenvalues of the Hamiltonian $\hat{H}_{23}$, namely 
\be
\hat{H}_{23}|E\rangle = E_{l}|E\rangle\,.
\label{cheigenket}
\ee
One then arrives at the energy eigenvalues or the Landau levels,
\be
E_l = \frac{|eB|}{2m}\hbar(2l + 1) = \hbar\omega(2l + 1)\,. 
\label{E23}
\ee 
All states with the same $l$ value regardless of $m$ values will thus have the same energy. Such states are $|l,-l,B\rangle\,,|l,-l+1,B\rangle\,,|l,-l+2,B\rangle\ldots$. These states form the irreducible representation of the symmetry group and the dimension of this representation is infinity. This, thus equals to the degeneracy of each energy level. In short, since $m\in[-l,\infty)$, there are infinite number of states with the same energy for a given $l$. One concludes that the Landau levels are infinitely degenerate. One can use (\ref{raiselowerkets}), the expression for the product $\hat{J}_{+}\hat{J}_{-}$ from (\ref{Cjm}) and the eigenvalues of the Casimir operator (\ref{Clow}) to determine $\alpha$ from the following amplitude,\footnote{We recall here the fact $\hat{J}^{\dagger}_{-} = \hat{J}_{+}$.}
\be
|\alpha|^{2} = \langle l\,,m\,,B|\hat{J}_{+}\hat{J}_{-}|l\,,m\,,B\rangle = -B\hbar(l + m)\,,\quad B < 0\,, 
\label{alpha}
\ee
which determines $\alpha$ up to a phase factor and therefore completely describing the action of the lowering operator $\hat{J}_{-}$ on kets $|l\,,m\,,B\rangle$,
\be
\hat{J}_{-}|l\,,m\,,B\rangle = \sqrt{-B\hbar(l + m)}|l\,,m-1\,,B\rangle\,,\quad B < 0\,.
\label{raiselowerkets2}
\ee
An immediate observation of the above result is when $m = -l$, the right hand side of (\ref{raiselowerkets2}) vanishes, as expected.
Other higher states can be generated from $|l\,,-l\,,B\rangle$ by successive applications of $\hat{J}_{+}$ according to,
\be
\hat{J}_{+}|l\,,m\,,B\rangle = \sqrt{-B\hbar(l + m + 1)}|l\,,m + 1\,,B\rangle\,,\quad B < 0\,,
\label{ketsgenenerateBneg}
\ee
which is obtained using (\ref{raiselowerkets}), (\ref{Cjp}) and (\ref{Clow}). In obtaining (\ref{ketsgenenerateBneg}), $\beta$ in (\ref{raiselowerkets}) is first determined (up to a phase factor) from,
\be
|\beta|^{2} = \langle l\,,m\,,B|\hat{J}_{-}\hat{J}_{+}|l\,,m\,,B\rangle = -B\hbar(l + m + 1)\,,\quad B < 0\,,
\label{beta}
\ee
which gives $\beta = \sqrt{-B\hbar(l + m +1)}$.

Next, we perform the calculation using the highest state: Acting $\hat{\bar{C}}$ on the highest state $|l\,,l\,,B\rangle$ as follows,
\be
\hat{\bar{C}}|l\,,l\,,B\rangle = 2\hat{J}_{-}\hat{J}_{+}|l\,,l\,,B\rangle + 2\hat{M}_{z}B|l\,,l\,,B\rangle + B\hbar\hat{\mathbb{I}}|l\,,l\,,B\rangle\,,
\label{Conhigh}
\ee 
and exploiting (\ref{jmax}), the following is obtained for the eigenvalues of the Casimir operator,
\be
C_l = B\hbar(2l + 1)\,,\quad B > 0\,.
\label{Chigh}
\ee
Using similar arguments as in the $B < 0$ case, one arrives at the following expression for the Landau levels:
\be
E_l = \frac{|eB|}{2m}\hbar(2l + 1) = \hbar\omega(2l + 1)\,. 
\label{E23b}
\ee 
We see that (\ref{E23}) and (\ref{E23b}) agrees with (\ref{eigenv}). As for the $B<0$ case, states with the same $l$ value regardless of $m$ values will thus have the same energy. Such states indeed form the irreducible representation of the symmetry group and is of infinite dimension. They are $|l,l,B\rangle\,,|l,l-1,B\rangle\,,|l,l-2,B\rangle\ldots$. Given that $m\in(-\infty,l]$, there are infinite number of states with the same energy for a given $l$. This again confirms the infinite degeneracy of Landau levels. One can now use (\ref{raiselowerkets}), the expression for the product $\hat{J}_{-}\hat{J}_{+}$ from (\ref{Cjp}) and the eigenvalues of the Casimir operator (\ref{Chigh}) to determine $\beta$ for $B>0$ case as follows,
\be
|\beta|^{2} = \langle l\,,m\,,B|\hat{J}_{-}\hat{J}_{+}|l\,,m\,,B\rangle = B\hbar(l - m)\,, 
\label{beta2}
\ee
which determines $\beta$ up to a phase factor and therefore completely describing the action of the raising operator $\hat{J}_{+}$ on kets $|l\,,m\,,B\rangle$,
\be
\hat{J}_{+}|l\,,m\,,B\rangle = \sqrt{B\hbar(l - m)}|l\,,m+1\,,B\rangle\,,\quad B > 0\,.
\label{raiselowerkets2b}
\ee
It also correctly yields the $m = l$ case, namely the vanishing of the right hand side of (\ref{raiselowerkets2b}).
Other lower states can be generated from $|l\,,l\,,B\rangle$ by successive applications of $\hat{J}_{-}$ according to,
\be
\hat{J}_{-}|l\,,m\,,B\rangle = \sqrt{B\hbar(l - m + 1)}|l\,,m - 1\,,B\rangle\,,\quad B > 0\,,
\label{ketsgenenerateBpos}
\ee
which is obtained using (\ref{raiselowerkets}), (\ref{Cjm}) and (\ref{Chigh}). In obtaining (\ref{ketsgenenerateBpos}), $\alpha$ in (\ref{raiselowerkets}) is first determined from,
\be
|\alpha|^{2} = \langle l\,,m\,,B|\hat{J}_{+}\hat{J}_{-}|l\,,m\,,B\rangle = B\hbar(l - m + 1)\,,\quad B > 0\,,
\label{beta}
\ee
which gives $\alpha = \sqrt{B\hbar(l - m + 1)}$.

\subsection{Landau levels in $|n_{+} - n_{-}\rangle$ basis}
Motivated by the connection between the $\bar{e}(2)$ algebra and the oscillator algebra, we will recalculate the Landau levels in an alternative basis.
From (\ref{e2barxoyogen}) and (\ref{numberpm}), it follows that $\hat{M}_{z}$ which is the $z$-component of the angular momentum operator, can be expressed in terms of the number operators, 
\be
\hat{M}_{z} = \hbar(\hat{N}_{+} - \hat{N}_{-})\,,
\label{Mznpnm}
\ee
where 
\be
\hat{N}_{+}|n_{+}\rangle = n_{+}|n_{+}\rangle\,,\quad \hat{N}_{-}|n_{-}\rangle = n_{-}|n_{-}\rangle\,,\quad \hat{N}_{\pm}|n_{+}-n_{-}\rangle = \pm n_{\pm}|n_{+}-n_{-}\rangle\,.
\label{numop}
\ee
$n_{+}$ and $n_{-}$ are the eigenvalues of the number operators $\hat{N}_{+}$ and $\hat{N}_{-}$ respectively. They can also be regarded as the quantum numbers associated with the eigenvalues of the Casimir operator and the Hamiltonian $\hat{H}_{23}$. As will be clear shortly, there is a correspondence between quantum numbers $\{n_{+}\,,n_{-}\}$ and $\{l\,,m\}$. Thus, an alternative basis formed by the set of state vectors $\{|n_{+} - n_{-}\rangle\}$ is indeed suitable for the calculation of Landau levels. Moreover, this set is also a common set of eigenkets for $\hat{\bar{C}}\,,\hat{M}_{z}$ and $\hat{B}$:
\be
\hat{\bar{C}}|n_{+}-n_{-}\rangle &=& C_{n_{+},n_{-}}|n_{+}-n_{-}\rangle\,,\quad \hat{M}_{z}|n_{+}-n_{-}\rangle = \hbar (n_{+}-n_{-})|n_{+}-n_{-}\rangle\,,\non\\ 
\hat{B}|n_{+}-n_{-}\rangle &=& B|n_{+}-n_{-}\rangle\,.
\label{npnmb}
\ee 
The commutation relations 
\be
\large[\hat{N}_{+}-\hat{N}_{-} , \hat{J}_{-}\large] = -\hat{J}_{-}\,,\quad \large[\hat{N}_{+}-\hat{N}_{-} , \hat{J}_{+}\large] = \hat{J}_{+}\,,
\label{jpmonnpm}
\ee
suggest that the lowering operator $\hat{J}_{-}$ lowers the eigenvalue of the $\hat{N}_{+}-\hat{N}_{-}$ by one. Conversely,
raising operator $\hat{J}_{+}$ raises the eigenvalue of the same operator by one. This can be written as,
\be
\hat{J}_{-}|n_{+}-n_{-}\rangle = \gamma|n_{+}-n_{-}-1\rangle\,,\quad \hat{J}_{+}|n_{+}-n_{-}\rangle = \delta|n_{+}-n_{-}+1\rangle\,. 
\label{jponnpmnm}
\ee 
The constants $\gamma$ and $\delta$ will be determined separately for cases $B<0$ and $B>0$. The next step is to identify the lowest and the highest states. The lowest state for a given $n_{-}$ is that for which $n_{+} = 0$ (This makes $n_{+}-n_{-}$ minimum for a given $n_{-}$.). Likewise, the highest state for a given $n_{+}$ is that for which $n_{-} = 0$ (This makes $n_{+}-n_{-}$ maximum for a given $n_{+}$.). Thus, the lowest state is denoted by $|-n_{-}\rangle$ and the highest state is denoted by $|n_{+}\rangle$. Consequently, analogous to (\ref{jmin}) and (\ref{jmax}), one can write the following to denote the action of the lowering and raising operators on the lowest and the highest states respectively,
\be
\hat{J}_{-}|-n_{-}\rangle = 0\,,\quad \hat{J}_{+}|n_{+}\rangle = 0\,.
\label{jpjmonnmpn}
\ee
Calculation of Landau levels using the lowest state: Using (\ref{Cjm}), we let $\hat{\bar{C}}$ to act on the lowest state $|-n_{-}\rangle$,
\be
\hat{\bar{C}}|-n_{-}\rangle = 2\hat{J}_{+}\hat{J}_{-}|-n_{-}\rangle + 2\hat{M}_{z}B|-n_{-}\rangle - B\hbar\hat{\mathbb{I}}|-n_{-}\rangle\,.
\label{connm}
\ee 
Exploiting the action of the lowering operator on the lowest state described by (\ref{jpjmonnmpn}) and using (\ref{npnmb}), one obtain the following for the eigenvalues of the Casimir operator,
\be
C_{n_{+},n_{-}} = -B\hbar(2n_{-} + 1)\,,\quad B < 0\,.
\label{cnm}
\ee
Re-evaluating the eigenvalues of the Casimir operator using the energy eigenket and (\ref{H23}) and equating these two expressions for $B<0$ case, one arrives at the following result for Landau levels in terms of $n_{-}$,
\be
E = \frac{|eB|}{2m}\hbar(2n_{-} + 1) = \hbar\omega(2n_{-} + 1)\,.
\label{Enm}
\ee
Note that although the above derivation utilizes the lowest state for which $n_{+} = 0$, all states such as $|1-n_{-}\rangle$,$|2-n_{-}\rangle$,$|3-n_{-}\rangle\ldots$, namely all states with the same $n_{-}$ regardless of $n_{+}$ values, would have the same energy value. Since $n_{+}$ can assume any positive integer $n_{+}\in[1,\infty)$, we have infinite number of states with the same energy value, which reflects infinite degeneracy of these levels. These states form the irreducible representation of the symmetry group and this representation has infinite dimension.

It is beneficial to describe the general results for the action of the raising and lowering operators on $|n_{+}-n_{-}\rangle$. Prior to this, calculations of $\gamma$ and $\delta$ must follow:  
One use (\ref{Cjm}), (\ref{npnmb}), (\ref{jponnpmnm}) and (\ref{cnm}) to determine $\gamma$ from the following amplitude,
\be
|\gamma|^{2} = \langle n_{+}-n_{-}|\hat{J}_{+}\hat{J}_{-}|n_{+}-n_{-}\rangle = -B\hbar n_{+}\,,\quad B < 0\,, 
\label{gamma}
\ee
which determines $\gamma$ up to a phase factor. This completely describes the action of the lowering operator $\hat{J}_{-}$ on kets $|n_{+}-n_{-}\rangle$,
\be
\hat{J}_{-}|n_{+}-n_{-}\rangle = \sqrt{-B\hbar n_{+}}|n_{+}-n_{-}-1\rangle\,,\quad B < 0\,.
\label{raiselowerkets2nm}
\ee
As a check, we notice the above result correctly yields the case for the lowest state, namely when $n_{+} = 0$, the right hand side of (\ref{raiselowerkets2nm}) vanishes. Other higher states e.g., $|-n_{-}+1\rangle\,,|-n_{-}+2\rangle$ etc., can be constructed from $|-n_{-}\rangle$ by successive applications of $\hat{J}_{+}$ according to,
\be
\hat{J}_{+}|n_{+}-n_{-}\rangle = \sqrt{-B\hbar(n_{+} + 1)}|n_{+} - n_{-} + 1\rangle\,,\quad B < 0\,,
\label{ketsgenenerateBnegnm}
\ee
which is again obtained using (\ref{Cjp}), (\ref{npnmb}), (\ref{jponnpmnm}) and (\ref{cnm}). In obtaining (\ref{ketsgenenerateBnegnm}), $\delta$ in (\ref{jponnpmnm}) was first determined (up to a phase factor) from,
\be
|\delta|^{2} = \langle n_{+}-n_{-}|\hat{J}_{-}\hat{J}_{+}|n_{+}-n_{-}\rangle = -B\hbar(n_{+} + 1)\,,\quad B < 0\,,
\label{delta}
\ee
which gives $\delta = \sqrt{-B\hbar(n_{+} +1)}$. 

Calculation of Landau levels using the highest state: (\ref{Cjp}) suggests,
\be
\hat{\bar{C}}|n_{+}\rangle = 2\hat{J}_{-}\hat{J}_{+}|n_{+}\rangle + 2\hat{M}_{z}B|n_{+}\rangle + B\hbar\hat{\mathbb{I}}|n_{+}\rangle\,.
\label{connm}
\ee 
Exploiting the action of the raising operator on the highest state (\ref{jpjmonnmpn}), one obtain the following for the eigenvalues of the Casimir operator,
\be
C_{n_{+},n_{-}} = B\hbar(2n_{+} + 1)\,,\quad B > 0\,.
\label{cnp}
\ee
Re-evaluating this eigenvalue using the energy eigenket and (\ref{H23}) and equating these two expressions for $B>0$ case, one arrives at the following for Landau levels in terms of $n_{+}$,
\be
E = \frac{|eB|}{2m}\hbar(2n_{+} + 1) = \hbar\omega(2n_{+} + 1)\,.
\label{Enp}
\ee
Although the derivation exploits the highest state for which $n_{-} = 0$, all states such as $|n_{+}-1\rangle$,$|n_{+}-2\rangle$,$|n_{+}-3\rangle\ldots$, namely all with the same $n_{+}$ regardless of $n_{-}$ values would have the same energy value. Since $n_{-}$ can assume any positive integer $n_{-}\in[1,\infty)$, we have infinite number of states with the same energy value, which reflects infinite degeneracy for each of these Landau levels. These states form the irreducible representation of the symmetry group with infinite dimension.

It is instructive to describe the general results for the action of the raising and lowering operators on $|n_{+}-n_{-}\rangle$ for $B>0$. We begin by determining $\gamma$ and $\delta$ for $B>0$. One uses (\ref{Cjp}), (\ref{npnmb}), (\ref{jponnpmnm}) and (\ref{cnp}) to calculate $\delta$,
\be
|\delta|^{2} = \langle n_{+}-n_{-}|\hat{J}_{-}\hat{J}_{+}|n_{+}-n_{-}\rangle = B\hbar n_{-}\,, 
\label{gamma}
\ee
which determines $\delta$ up to a phase factor. This completely describes the action of the raising operator $\hat{J}_{+}$ on kets $|n_{+}-n_{-}\rangle$,
\be
\hat{J}_{+}|n_{+}-n_{-}\rangle = \sqrt{B\hbar n_{-}}|n_{+}-n_{-}+1\rangle\,,\quad B > 0\,.
\label{raiselowerkets2np}
\ee
The above result correctly yields the case for the highest state for which $n_{-} = 0$, when the right hand side of (\ref{raiselowerkets2np}) vanishes.
Other lower states e.g., $|n_{+}-1\rangle\,,|n_{+}-2\rangle$ etc., can be constructed from $|n_{+}\rangle$ by successive applications of $\hat{J}_{-}$ according to,
\be
\hat{J}_{-}|n_{+}-n_{-}\rangle = \sqrt{B\hbar(n_{-} + 1)}|n_{+} - n_{-} - 1\rangle\,,\quad B > 0\,.
\label{ketsgenenerateBnegnp}
\ee
In obtaining (\ref{ketsgenenerateBnegnp}), $\gamma$ is determined from,
\be
|\gamma|^{2} = \langle n_{+}-n_{-}|\hat{J}_{+}\hat{J}_{-}|n_{+}-n_{-}\rangle = B\hbar(n_{-} + 1)\,,\quad B > 0\,,
\label{delta}
\ee
which then yields $\gamma$ up to a phase factor. 

An interesting correspondence emerges: Comparison of (\ref{raiselowerkets2}) with (\ref{raiselowerkets2nm}) (or (\ref{ketsgenenerateBneg}) with (\ref{ketsgenenerateBnegnm})) suggests $n_{+}\to l+m$. This supports the fact that the lowest state expressed as $|l,-l,B \rangle$ (where $m = -l$) is indeed equivalent to $|- n_{-}\rangle$ (where $n_{+} = 0$) as pointed out above.  Similarly comparing (\ref{raiselowerkets2b}) with (\ref{raiselowerkets2np}) (or (\ref{ketsgenenerateBpos}) with (\ref{ketsgenenerateBnegnp})) suggests $n_{-}\to l-m$ that supports the fact that the highest state expressed as $|l,l,B \rangle$ (where $m = l$) is indeed equivalent to $|n_{+}\rangle$ (where $n_{-} = 0$). Such a correspondence lies at the heart of the Schwinger's oscillator model of angular momentum that highlights the connection between the oscillator algebra and the $so(3)$ algebra (see \cite{SakuraiNep}). We find it interesting that the same correspondence between $\{l\,,m\}$ and $\{n_{+}\,,n_{-}\}$ exists here where a connection is found between oscillator algebra and the $\bar{e}(2)$ algebra instead.

\section{Discussion}
Landau levels are rederived from purely symmetry considerations. In particular, the Casimir operator of planar Landau system's symmetry group $\bar{E}(2)$, is used for the purpose. The calculation is presented separately in two sets of basis, namely basis formed by eigenkets $|l\,,m\,,B\rangle$ and $|n_{+}-n_{-}\rangle$ respectively. They are subsequently shown to be related to each other. The correspondence between the two sets of quantum numbers $\{l\,,m\}$ and $\{n_{+}\,,n_{-}\}$ is laid out. Such a relation exists within the framework of Schwinger's oscillator model of angular momentum, in which connections are made between the oscillator algebra and $so(3)$ algebra. Interestingly, a similar relation is found to exist here between the oscillator algebra and the $\bar{e}(2)$ algebra. This fascilitated the calculations of the Landau levels in the $|n_{+}-n_{-}\rangle$ basis. The well-known infinite degeneracy of Landau levels that results from the existence of conserved observables that reflects the $\bar{E}(2)$ symmetry is discussed and revisited in terms of irreducible representations of the symmetry group.   

We believe that beginning graduate students will benefit greatly from this coherent approach of the subject based on inherent symmetry of the system. In addition, given the importance of symmetry in quantum mechanics, we find that it will be advantageous for students to be exposed to calculation techniques that relies on the framework of group theory which is often insufficiently stressed in standard texts. It is our hope that this paper will serve as a good starting point for students to explore the literature on the subject going forward.

\end{document}